\documentclass[aps,pre,showpacs,twocolumn]{revtex4-1}

\usepackage{amsfonts}
\usepackage[dvips]{graphicx}

\begin{document}

\title{Cluster Algorithm Renormalization Group Study of\\Universal Fluctuations in the 2D Ising Model}
\author{G. Palma}
\email{guillermo.palma@usach.cl}
\author{D. Zambrano}
\affiliation{Departamento de F\'{\i}sica, Universidad de Santiago de Chile,\\
Casilla 307, Santiago 2, Chile.}

\begin{abstract}

In this paper we propose a novel method to study critical systems
numerically by a combined collective-mode algorithm and
Renormalization Group on the lattice. This method is an improved
version of MCRG in the sense that it has all the advantages of
cluster algorithms. As an application we considered the 2D Ising
model and studied wether scale invariance or universality are
possible underlying mechanisms responsible for the approximate
\textquotedblleft universal fluctuations" close to a so-called bulk
temperature $T^\ast(L)$. \textquotedblleft Universal fluctuations"
was first proposed in \cite{BHP} and stated that the probability
density function of a global quantity for very dissimilar systems,
like a confined turbulent flow and a 2D magnetic system, properly
normalized to the first two moments, becomes similar to the
\textquotedblleft universal distribution", originally obtained for
the magnetization in the 2D XY model in the low temperature region.
The results for the critical exponents and the renormalization group
flow of the probability density function are very accurate and show
no evidence to support that the approximate common shape of the PDF
should be related to both scale invariance or universal behavior.

\end{abstract}


\maketitle

\section{Introduction}

Critical phenomena are present in a large number of quite different
physical systems: super-fluid Helium three, low temperature
super-conductors, ferromagnetic- paramagnetic systems, turbulent
fluids, plasmas,  polymers, among many others. Nevertheless, an
important common feature of these systems is the scale independent
fluctuations at the critical temperature: although the underlying
inter-molecular forces, responsible for the existence of phase
transitions, have a well-defined length scale, the structures they
give rise do not. This leads, very close to the critical temperature
(scaling region), to a power-law behavior of the physical
quantities, which is a fundamental feature of universality
\cite{BDFN}.

The main challenge of the theory of critical phenomena is to explain
how dissimilar systems exhibit the same critical behavior.
Renormalization Group (RG) Theory developed by Wilson and Kogut
\cite{WK} provides a consistent framework to understand the
existence of equivalence classes of critical systems. On the
lattice, a very cunning method of applying a RG analysis to Monte
Carlo simulations of general systems was first proposed in
\cite{MCRG}.

Nevertheless, the numerical simulation of critical systems has a
serious limitation due to the critical slowing down effect. Indeed,
as a critical system approaches the critical temperature, the
decorrelation time diverges with the power of the correlation length
of the system $\xi$ to the dynamical critical exponent $z$: $\tau
\sim \xi^z$, where $\xi$ is approximately 2 for local-flip
algorithms like the Metropolis algorithm. In order to beat or at
least reduce this effect, a cluster algorithm was first developed in
\cite{SWW}.

In this paper, we develop a self-consistent method along the line
proposed by Swendsen in \cite{MCRG}, which will be explained in
section II, to perform a Lattice Renormalization Group (RG) analysis
of the probability density function (PDF) as a function of the
magnetization, in the 2D Ising model. His proposal amounts to
compute by a direct Monte Carlo simulation of the fundamental
Hamiltonian, a sequence of approximations to the linearized RG
matrix $T_{\alpha\beta}{}^\ast$. From its eigenvalues the critical
exponents can be obtained in a direct way. In order to reduce
critical slowing down, we used instead of a local (Metropolis-type)
algorithm the collective-mode algorithm developed in \cite{UW} to
simulate the fundamental Hamiltonian. We call this method Cluster
Algorithm Renormalization Group (CARG) \cite{lat08}.

Our physical motivation is to study, in the context of the 2D Ising
model, whether scale invariance and universality are the underlying
mechanism which could give rise to the approximate \textquotedblleft
universal curve". This phenomenon link a large class of dissimilar
systems defined on different dimensions and including
non-equilibrium systems, in which the PDF of a global quantity -like
the power consumption in a confined turbulent system or the
magnetization in a finite ferromagnetic system- properly normalized
to the first two moments, is described by a single curve \cite{PRL}.
This claim is not free of controversy, and in the last years it has
been the central task of several publications
\cite{ZT}-\cite{CHAPMAN}. This \textquotedblleft universal curve"
was shown to correspond to the PDF of the 2D-XY model in the zero
temperature limit \cite{MPV}.

The paper is organized as follows: in section II the CARG method is
explained for a general critical system. In the third section we
explain the concept of approximate \textquotedblleft universal
fluctuations" of the PDF of the magnetization for the 2D Ising
model. In section IV the critical exponents $\nu$ and $\eta$ are
computed using fundamental lattices of lattice sizes $L=45$, $L=64$
and $L=108$, performing two and three RG steps, to check the
accuracy of the method. It is shown that relative errors of the
critical exponents increase monotonically with the departure from
the critical temperature. Finally, the RG flow of the PDF itself is
computed and it is shown that when conveniently normalized, it is
invariant. The underlying reason for this invariance is discussed in
connection with the so-called \textquotedblleft generalized
universality" conjecture. The results are discussed and conclusions
are formulated in the last section.

\section{The CARG Method}

To describe the method, we first introduce the notation. We consider
a model defined in a square lattice of lattice spacing $a$, and
linear size $L$, with periodic boundary conditions (p.b.c.). The
spin variables $\sigma_i$ are defined on each site $i$ of the
lattice and the Hamiltonian has the form
\begin{equation}
H=\sum_\alpha K_\alpha S_\alpha
\label{def}
\end{equation}
where each $S_\alpha$ is some kind of combination of the spin
variables, with the only requirement of translation invariance
subject to p.b.c.. For our study we will consider the 2D Ising
model, including up to three even interactions (nearest-neighbor,
second-neighbor, and four spin) and one odd interaction (a weak
magnetic field).

The aim of RG theory is to study the critical properties of a model.
The critical exponents for example, can be obtained from the
linearized RG transformation matrix $T_{\alpha\beta}{}^\ast$,
defined in eqn. (\ref{LRGT}), by computing its eigenvalues.
$T_{\alpha\beta}{}^\ast$ can be obtained numerically from the
coupled equations,

\begin{equation}
\frac{\partial\langle~S_\gamma{}^{(n)}~\rangle}{\partial
K_\beta{}^{(n-1)}}= \sum_\alpha\frac{\partial
K_\alpha{}^{(n)}}{\partial
K_\beta{}^{(n-1)}}\frac{\partial\langle~S_\gamma{}^{(n)}~\rangle}{\partial
K_\alpha{}^{(n)}} \label{RGT}
\end{equation}

Here, the super-indices $(n-1)$ and $(n)$ denote the original and
the renormalized quantity respectively, after one RG transformation,
and the thermal average $\langle \ F(\phi) \ \rangle$ is defined by
the usual formula

\begin{equation}
\langle \ F (\sigma) \ \rangle=\frac{1}{Z}\sum_{\{ \sigma\}} \ F \
({\sigma} )
 \exp{(-H(\sigma))}
\end{equation}

The left and right quantities appearing in eqn. (\ref{RGT}) can be
obtained through the identities

\begin{equation}
\frac{\partial\langle~S_\gamma{}^{(n)}~\rangle}{\partial
K_\beta{}^{(n-1)}}=-\langle~S_\gamma{}^{(n)}~S_\beta{}^{(n-1)}~\rangle+\langle~S_\gamma{}^{(n)}~\rangle~\langle~S_\beta{}^{(n-1)}~\rangle
\label{RGT1}
\end{equation}

\begin{equation}
\frac{\partial\langle~S_\gamma{}^{(n)}~\rangle}{\partial
K_\alpha{}^{(n)}}=-\langle~S_\gamma{}^{(n)}~S_\alpha{}^{(n)}~\rangle+\langle~S_\gamma{}^{(n)}~\rangle~\langle~S_\alpha{}^{(n)}~\rangle
\label{RGT2}
\end{equation}

Due to the definition of the renormalized Halmiltonian, the thermal
expectation of any function $F$ of the variables $\sigma{}^{(n)} $
yield the same value, whether one evaluates it using $H{}^{(n)}$ or
$H{}^{(n-1)}$, i.e.:

\begin{eqnarray}
\frac{1}{Z{}^{(n)}} \sum_{\{ \sigma{}^{(n)}\}} &&
 \hspace{-.7 cm} \ F( \sigma{}^{(n)}) \ e^{-H{}^{(n)}( \sigma{}^{(n)})} =
\\ && \frac{1}{Z{}^{(n-1)}}\sum_{\{ \sigma{}^{(n-1)}\}} \ F(
\sigma{}^{(n-1)}) \ e^{-H{}^{(n-1)}(\sigma{}^{(n-1)})} \nonumber
\end{eqnarray}

Therefore, the quantities appearing in equations (\ref{RGT1}) and
(\ref{RGT2}) can be computed directly by a numerical simulation of
the original Hamiltonian on the fundamental lattice. Because the
simulations must be performed close to the critical point of the
system, the configurations produced by a local-update algorithm are
not statistically independent, which leads to inaccurate thermal
averages. This phenomenon is called in the literature critical
slowing down \cite{BDFN}. In order to avoid this effect we use the
collective-mode algorithm developed by Wolff \cite{UW} for the Ising
model.

The expectation values appearing in equations (\ref{RGT1}) and
(\ref{RGT2}) should be computed by using a simulation with a cluster
algorithm on the fundamental lattice. These numerical values
inserted into eqn. (\ref{RGT}) lead to coupled algebraic equations,
which allow to obtain a sequence of approximations to the linearized
RG transformation $T_{\alpha\beta}{}^\ast$, as the renormalization
transformation is iterated to the fixed point Hamiltonian $H^\ast$,
defined in the vicinity of the fixed point by
\begin{equation}
K_\alpha{}^{(n+1)}-K_\alpha{}^\ast=\sum_\beta
T_{\alpha\beta}{}^\ast(~K_\beta{}^{(n)}-K_\beta{}^\ast~) \label{T*}
\end{equation}
where the linearized RG transformation is defined by
\begin{equation}
T_{\alpha\beta}{}^\ast=\left[~\frac{\partial
K_\alpha{}^{(n)}}{\partial K_\beta{}^{(n-1)}}~\right]_{H^\ast}
\label{LRGT}
\end{equation}

To evaluate $T_{\alpha\beta}{}^\ast$ one in principle needs a
\textquotedblleft linear region" close to the critical temperature,
where its derivatives are essentially constant. This region can be
found by using FSS analysis of the lattice shifted critical
temperature \cite{lat08}, and in the case of the 2D Ising model, by
using the Binder Cumulant (see section IV).

Finally, the critical exponents are obtained from the eigenvalues of
$T_{\alpha\beta}{}^\ast$ from eqn. (\ref{LRGT}) in the standard way.
For the 2D Ising model they are given by the fundamental relations
$\nu=\ln s/\ln\lambda_1{}^{e}$ and $\eta=d+2-2\ln\lambda_1{}^{o}/\ln
b$, where $\lambda_1{}^{e(o)}$ is the largest even (odd) eigenvalue.

\section{Universal fluctuations in the 2D Ising Model}

The 2 dimensional Ising model is well known and it is defined by the
Hamiltonian

\begin{equation}
H(\sigma )=\frac{J}{k_BT} \sum_{<i,j>} \sigma_i ~ \sigma_j
\label{Ising}
\end{equation}
where $\sigma_i$ is the spin variable defined on the lattice site i,
$J>0$ is the ferromagnetic constant, $k_B$ is the Boltzmann's
constant and $\sum_{<i,j>}$ stands for sum over nearest neighbors
$<i,j>$. We use a system of units where Boltzmann's constant is set
equal to unity throughout the paper and identify $T$ with the
reduced temperature $T/J$. It is well known that for an infinite
square lattice ($L\to \infty$), this model has a second order phase
transition at the critical temperature $T_{c}$ defined by
$\sinh(1/T_{c})=1$ or equivalently $T_{c} = 2/ln(1+\sqrt(2))$, which
was first computed by Onsager \cite{Onsager}.

Universality of Rare Fluctuations in Turbulence and Critical
Phenomena, as it was first proposed in \cite{BHP}, linked two quite
different physical systems: a confined turbulent flow and the finite
volume 2D XY model in the low temperature regime. The (generalized)
universality of both systems was based on the collapse of the
probability distribution functions of the corresponding global
quantities, the power consumption in the turbulent experiment and
the magnetization in the critical system. The PDFs were conveniently
normalized to their first two moments, and felt onto a common curve,
which was called \textquotedblleft universal fluctuations".

A further generalization of the phenomenon described above was
performed in \cite{PRL}, including non-equilibrium models like the
auto igniting forest fire model and the Bak-Tang-Wiesenfeld sandpile
model, and the equilibrium 2D Ising model at temperature
$T^\ast(L)$, which they called \textquotedblleft bulk temperature".

Nevertheless, using the 2D and 3D Ising model Zheng et al. \cite{ZT}
showed a dependence of the \textquotedblleft universal fluctuations"
on the equivalence class of the model. In a high precision MC
simulation of the finite volume 2D XY-model, a slight but systematic
dependence of the \textquotedblleft universal fluctuations" on the
system temperature was first suggested \cite{PML}. This claim was
proved analytically in \cite{MPV}. Moreover, by computing the
skewness and the curtosis in the harmonic 2D XY-model (the second
and third moments of the PDF) the temperature dependence of the PDF
was confirmed numerically and analytically in refs. \cite{BB} and
\cite{P} respectively.

The precise physical definition of $T^\ast(L)$ is still in progress.
In more recent papers, $T^\ast(L)$ has been linked to intermittency
of the magnetization \cite{2DIM}.

To address this issue we used the CARG method described in the above
section and we have simulated the 2D Ising model in a square lattice
of lattice sizes $L=45$, $L=64$ and $L=108$, with periodic boundary
conditions, in a range of temperatures within the bulk $T^\ast(L)$
and the lattice shifted critical temperature $T_c(L)$ which can be
defined, for example, as the temperature at which the magnetic
susceptibility has a peak.

\begin{figure}[th]
\centering
\includegraphics[scale=0.5]{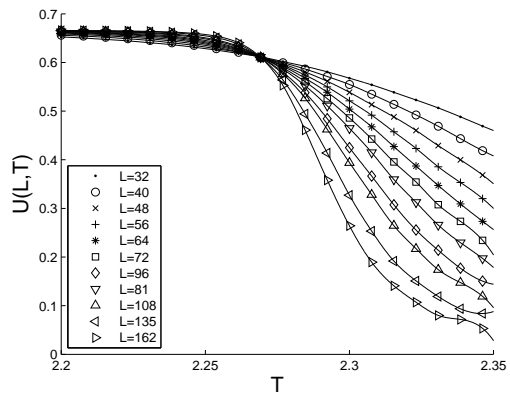}
\caption{The Binder Cumulant is shown for different lattice sizes.
From their interception the critical temperature for infinite
volume, $T_c(\infty)=2.26903\pm0.00059$, is obtained. It differs in
0.7 per one thousand from Onsager's exact result.} \label{fig.3}
\end{figure}

\section{Numerical Results}

As it was mentioned in section II, and estimation for the critical
temperature for the 2D Ising model can be obtained by using the
Binder Cumulant \cite{BINDER}:

\begin{equation}
u(L,T)=1-\frac{1}{3}\frac{\langle~M{}^{4}~\rangle}{\langle~M{}^{2}~\rangle
^{2}}. \label{BC}
\end{equation}
where $ \langle M{} \rangle$ is the averaged magnetization. It
follows that $u(L,T)$ depends on the system size and temperature
through the ratio of the averaged fourth power to the averaged
square of the magnetization. Nevertheless, from finite size scaling
analysis, close to the critical temperature it is independent of the
lattice size, and therefore, the curves representing the Binder
cumulant corresponding to different lattice sizes must collapse onto
one point. This behavior is displayed in Fig. \ref{fig.3}. The value
obtained agrees with Onsager's critical temperature within an error
of 0.7 per one thousand.

Now we use CARG method to compute the critical exponents $\nu$ and
$\eta$, associated to the Ising model. In order to compare with the
results obtained in \cite{MCRG}, we use a lattice of lattice size
$L=45$ to compute the quantities appearing in the RG equations
(\ref{RGT1})-(\ref{RGT2}) close to the critical temperature,
obtained as explained above, and perform two RG steps. Using the
Wolff algorithm \cite{UW}, $10^5$ sweeps were used to thermalize the
system and $10^6$ configurations were used to compute thermal
averages.

For the RG analysis up to three even interactions (nearest-neighbor,
second-neighbor and four spin or plaquette) and one odd interaction
(magnetic field) were considered. In order to compare our results
with the corresponding ones obtained by \cite{MCRG}, we display in
table \ref{tab.1} the results of a lattice of lattice size $L=45$,
where the scale factor $b=3$ was used. After two RG steps, the
values compare quite well. Compared to their exact values of the
infinite volume limit, they agree up to $0.25\%$ for $\nu$ and
$0.52\%$ for $\eta$.

\begin{table}[th]
\caption{The critical exponents obtained by the CARG method [PZ] and
by Swendsen's method [S] for $L=45$ are displayed.} \label{tab.1}
\begin{center}
\begin{tabular}{|c||c|c|c|c|}
\hline
\hline RG Step     & $\lambda_1{}^e$ & $\lambda_1{}^o$ & $\nu$     & $\eta$   \\
\hline $1_{[S]}$   & $2.887$         & $7.712$         & $1.036$   & $0.2812$ \\
\hline $1_{[PZ]}$  & $2.8900$        & $7.7408$        & $1.0352$  & $0.2744$ \\
\hline $2_{[S]}$   & $3.006$         & $7.835$         & $0.998$   & $0.2524$ \\
\hline $2_{[PZ]}$  & $3.0083$        & $7.8508$        & $0.9975$  & $0.2487$ \\
\hline Exact       & $3$             & $7.8452$        & $1$       & $0.250$  \\
\hline \hline
\end{tabular}
\end{center}
\end{table}

\medskip

In table \ref{tab.2} we display the results for the critical
exponents $\nu$ and $\eta$, for a lattice of lattice size $L=64$.
The scale factor $b=2$ was used. The results are even better than
with the smaller lattice, with errors of order less than 0.13\%. The
reason for this behavior is physically intuitive: An important
feature of the RG method is that the smallest system considered
should be still large compared to the range of the fixed-point
Hamiltonian, so that any significant truncation should be avoided.
The validity of this requirement is improved with the larger lattice
sizes.

\begin{table}[th]
\caption{Critical exponents for $L=64$ and $T=2.259$.} \label{tab.2}
\begin{center}
\begin{tabular}{|c||c|c|c|c|}
\hline
\hline RG Step & $\lambda_1{}^e$ & $\lambda_1{}^o$ & $\nu$    & $\eta$   \\
\hline $1$     & $1.9586$        & $3.6856$        & $1.0311$ & $0.2362$ \\
\hline $2$     & $1.9986$        & $3.6676$        & $1.0010$ & $0.2503$ \\
\hline Exact   & $2$             & $3.6680$        & $1$      & $0.250$  \\
\hline \hline
\end{tabular}
\end{center}
\end{table}

In spite of the good accuracy obtained for the critical exponents
performing only two RG steps, we iterated the RG transformations to
three RG steps in order to compare our results with the ones
obtained in reference \cite{MCRG2}. The comparison is displayed in
table III. Both methods achieve a remarkable accuracy, compared to
the exact results. The main difference is related to the
computational time needed to obtained these values. Indeed, it
turned out that the CARG method is faster than Swendsen's method by
a factor which increases monotonically with the lattice size, from
six for $L=64$ until 10 for $L=162$.

\begin{table}[th]
\caption{The critical exponents obtained by CARG method [PZ] and
those obtained by Swendsen \cite{MCRG2} for $L=108$.} \label{tab.3}
\begin{center}
\begin{tabular}{|c||c|c|c|c|}
\hline
\hline RG Step     & $\lambda_1{}^e$ & $\lambda_1{}^o$ & $\nu$     & $\eta$   \\
\hline $1_{[S]}$   & $2.852$         & $7.705$         & $1.048$   & $0.2828$ \\
\hline $1_{[PZ]}$  & $2.8635$        & $7.7062$        & $1.0443$  & $0.2825$ \\
\hline $2_{[S]}$   & $3.021$         & $7.828$         & $0.994$   & $0.2540$ \\
\hline $2_{[PZ]}$  & $3.0027$        & $7.8269$        & $0.9992$  & $0.2542$ \\
\hline $3_{[S]}$   & $3.007$         & $7.831$         & $0.998$   & $0.2534$ \\
\hline $3_{[PZ]}$  & $3.0013$        & $7.8361$        & $0.9996$  & $0.2521$ \\
\hline Exact       & $3$             & $7.8452$        & $1$       & $0.250$  \\
\hline \hline
\end{tabular}
\end{center}
\end{table}

We want now to address the question whether the model displays or
not a critical behavior at the bulk temperature $T^\ast(L)$, which
would lead to scale invariance of the system. By definition, at
$T^\ast(L)$ the PDF of the magnetization has a similar form to the
distribution originally obtained for the magnetization in the 2D
XY-model in the zero temperature limit \cite{MPV}. The scale
invariance is usually expressed as a power law behavior of the
physical quantities involved and therefore we use the CARG method to
look for critical exponents in the whole inertial range
[$T^\ast(L)$, $T_c(L)$] \cite{PRL}.

In reference \cite{2DIM}, $T^\ast(L)$ was defined as the temperature
for which the skewness ( the third normalized moment of the PDF of
the magnetization ) is equal to that for the 2D XY model in the low
temperature region. According to this definition, and for a lattice
of lattice size $L=64$ the numerical value $T^\ast(L)=2.11$ was
found. In this article it was argued that at $T^\ast(L)$ there are
(strong) correlations on all scales up to a length of the order of
the system size $L$, which is an important feature of critical
behavior. Furthermore, the fluctuations of the magnetization were
linked to intermittency. To address this issue we have simulated a
lattice with $L=64$ as well.

\begin{figure}[th]
\centering
\includegraphics[scale=0.5]{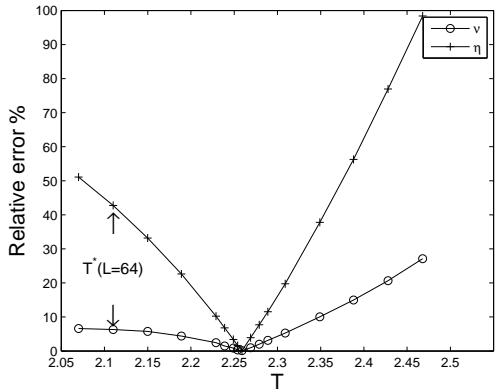}
\caption{Relative errors for the critical exponents $\eta$ and $\nu$
obtained in the second renormalization step with lattice size $L=64$
and $NAV=10^6$. The minimum error occur at $T_c(64)=2.259$ and the
critical exponents for this temperature are shown in table
\ref{tab.2}.} \label{fig.4}
\end{figure}

In Fig. \ref{fig.4}, the relative errors between the exact values of
the infinite volume critical exponents $\nu$ and $\eta$ and the
corresponding numerical result are displayed as a function of the
system temperature. They are expressed in percents and from this
figure we conclude that the relative errors increase monotonically
as the temperature moves away from $T_c(L)$, which includes the
particular value $T^*(L)$, at which the relative errors are large.
It is therefore rather unjustified to expect scale invariance of the
physical quantities close to $T^*(L)$.

\medskip

Now we study a possible scaling behavior of the order parameter -the
magnetization- in the vicinity of both the bulk and the shifted
critical temperatures ($T^*(L)$ and $T_c(L)$)respectively. It is
well known that the magnetization near the critical temperature
behaves as a power law of the kind
\begin{equation}
\left\langle M \right \rangle \sim\tau^\beta
\end{equation}
where the reduced temperature $\tau$ is given by
\begin{equation}
\tau=\frac{|T-T_c(L)|}{T_c(L)}
\end{equation}
and $\beta$ is the critical exponent. A direct numerical computation
of this exponent is rather subtle because of the reflection symmetry
of the Hamiltonian (\ref{Ising}), which leads in numerical
simulations close to the phase transition to the use of the absolute
value of the magnetization instead of the magnetization itself. One
can instead compute $\beta$ by using the scaling relation
$\beta=\nu(d+\eta-2)/2$.

In ref. \cite{2DIM} the stochastic evolution of the magnetization
was computed and it was argued that the system displays
\textquotedblleft larger fluctuations" at $T^*(L)$ than at $T_c(L)$.
This apparent phenomenon has been called \textquotedblleft
intermittency". Nevertheless, and according to the discussion
associated to fig. \ref{fig.4}, no underlying scale invariance is
responsible for this phenomenon. This apparent behavior is rather
related to the fact that, in order to avoid metastable states in
numerical simulations, the observable used in MC simulations is the
absolute value of the magnetization instead of the magnetization
itself. This quantity is bounded from below and therefore at
$T_c(L)$ its fluctuations appear suppressed by a factor of order 2
compared to the corresponding at $T^*(L)$, where this lower bound
plays no role as the magnetization does not vanishes in its
neighborhood.

\subsection*{RG-flow of the Probability Density Function}

\begin{figure}[th]
\centering
\includegraphics[scale=0.5]{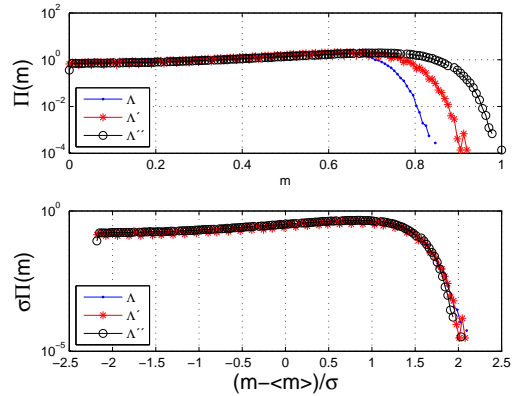}
\caption{PDFs (raw and normalized data) for the fundamental lattice
($L=64$) at temperature $T_c(L)=2.3008$ and for the two coarse
grained lattices ($L=32$ and $L=16$).} \label{fig.5}
\end{figure}

In this subsection, we proceed to study of the PDF by using the CARG
method. In particular we have computed the RG flow of the PDF
starting on a lattice of lattice size $L=64$, which we called
fundamental lattice, which we denote by $\Lambda$. We have performed
two RG transformations. Two further lattices were defined with
lattice sizes $L=32$ and $L=16$, which are denoted by $\Lambda'$ and
$\Lambda''$ respectively. In fig. \ref{fig.5} the PDF is displayed
for the three lattices. It follows that, properly normalized to the
first two moments, the PDF remains invariant along a renormalized
trajectory, see second graphic in fig. \ref{fig.5}. For the present
computation we used Swendsen-Wang's cluster algorithm \cite{SWW} to
update the fundamental lattice. We used the block spin parameter
$b=2$ and $T_c(L=64)=2.3008$ for the lattice shifted critical
temperature.

\medskip

The invariance of the PDF under RG transformations can be understood
by observing that it can be written as a Fourier transform of a
partition function of an auxiliary theory, which differs from the
original theory by a dimension zero perturbation, with a very small
imaginary coefficient. This was first pointed out in \cite{MPV}.
From the RG theory we know that partition functions are invariant
under RG transformations.

\begin{figure}[th]
\centering
\includegraphics[scale=0.5]{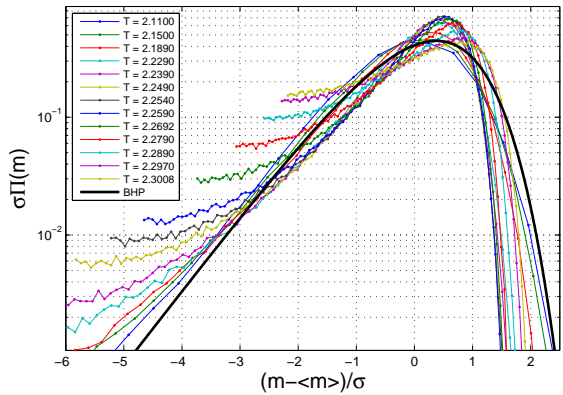}
\caption{The PDF is displayed for temperatures in the
\textquotedblleft inertial range" and for the lattice size $L=64$.}
\label{fig.8}
\end{figure}

Finally we plot in figure \ref{fig.8} the PDF in the whole
\textquotedblleft inertial range" of temperatures. The curves depend
slightly on the temperature close to $T^*(L)$, but this particular
value, as we have argued, is not related to scale invariance or
power law behavior of the system. Moreover, different curves
normalized to the first two moments are difficult to distinguish
from one another, close to the maximum. This has been pointed out in
\cite{CHAPMAN}, where the PDF at the particular value $T^*(L)$ was
fitted by a Gumbel distribution, which plays an important role in
Statistics of Extremes.

\medskip

\section{Conclusions}

In this paper we have introduced a very accurate method, which we
call CARG method, to study critical systems close to a critical
point. Compared to the original method proposed by Swendsen, our
method is faster by a factor which grows linearly with the lattice
size from 6 for $L=64$ until 10 for $L=162$, due to the use of
cluster algorithms to simulate the fundamental Hamiltonian. This
advantage allows to simulate larger lattices in reasonable
computational times, and to improve the accuracy of the results, as
shown in table III.

We illustrated the method by using the 2D Ising model defined in a
square lattice of lattice size $L$, and we have shown how to obtain
very accurate values for the critical exponents without the previous
knowledge of the critical temperature. We used further the 2D Ising
model to study scale invariance and universality as the underlying
mechanism which could give rise to the approximate generalized
universal behavior of fluctuations. We computed the probability
density function (PDF) of the magnetization and its RG trajectory
close to the lattice shifted critical temperature.

Critical behavior is associated to scale invariance, which is
commonly represented as a power law behavior: the physical
quantities are described close enough to a critical point, by
critical exponents, which characterize the equivalence class the
system belongs to. In the special case of the 2D Ising model, from
Fig. \ref{fig.4} we conclude that, at the bulk temperature
$T^\ast(L)$, neither scale invariance nor universal behavior is
really present. Therefore the approximate collapse of the PDF onto
the \textquotedblleft universal distribution" seems not to be
related to critical behavior but rather to a numerical phenomenon
associated to approximated scaling relations and extreme statistics,
as proposed in \cite{CHAPMAN}, and to the constraint character of
the global quantity used to compute the PDF.

\acknowledgments This work was partially supported by FONDECYT Nº
1050266 and DICYT.

\end{document}